\documentclass[12pt]{article}
\usepackage{a4wide,axodraw}

\begin{document}

\title{
\vskip-3cm{\baselineskip14pt
\centerline{\normalsize DESY 05-213\hfill ISSN 0418-9833}
\centerline{\normalsize hep-ph/0510235\hfill}
\centerline{\normalsize October 2005\hfill}}
\vskip1.5cm
Two-loop sunset diagrams with three massive lines}

\author{B.A.~Kniehl${}^a$, A.V.~Kotikov${}^{a,b}$, A.~Onishchenko${}^{a,c}$,
O.~Veretin${}^{a,d}$\\
\\
${}^a$ {\normalsize\it II Institut f\"ur Theoretische Physik,
Universit\"at Hamburg},\\
{\normalsize\it 22761 Hamburg, Germany}\\
${}^b$ {\normalsize\it Bogolyubov Laboratory for Theoretical Physics, JINR,}\\
{\normalsize\it 141980 Dubna (Moscow region), Russia}\\
${}^c$ {\normalsize\it Theoretical Physics Department,
Petersburg Nuclear Physics Institute,}\\
{\normalsize\it Orlova Roscha, 188300 Gatchina, Russia}\\
${}^d$ {\normalsize\it Petrozavodsk State University,}\\
{\normalsize\it 185910 Petrozavodsk, Karelia, Russia}}

\date{}

\maketitle

\begin{abstract}
In this paper, we consider the two-loop sunset diagram with two different
masses, $m$ and $M$, at spacelike virtuality $q^2=-m^2$.
We find explicit representations for the master integrals and an analytic
result through $O(\varepsilon)$ in $d=4-2\varepsilon$ space-time dimensions
for the case of equal masses, $m=M$.

\medskip

\end{abstract}

\newpage

\section{Introduction}

The two-loop self-energy diagram with three lines, the so-called sunset
diagram, plays a very important role in the evaluation of higher-order
corrections in quantum electrodynamics (QED), quantum chromodynamics (QCD),
and the electroweak theory.
In fact, it enters almost every calculation of such kind.
Therefore, it has been an object of intensive investigations in the literature
(see, e.g., Ref.~\cite{sunsetpub}).
In the cases when one or two masses are zero, analytical results in terms of
polylogarithmic functions are available
\cite{Kotikov:1990kg,Fleischer:1998dw} even at arbitrary external momentum
$q$.

A much more difficult situation arises when all three lines are massive. 
In this case, the result cannot be expressed in terms of polylogarithms, and
more involved functions must be considered \cite{Berends:1993ee}.
In particular, the case when all three masses are equal, $m_1=m_2=m_3=m$, is
of great importance.
Calculations for such kinematics were carried out using the
dispersion-relation representation \cite{Davydychev:2003cw}, the
differential-equation approach, and several expansions in the variable
$q^2/m^2$ \cite{Broadhurst:1993mw}.
Recently, a detailed analysis of the differential-equation approach, together
with solutions and analytical continuations, was presented
\cite{Laporta:2004rb}.
At singular points, including zero invariant mass ($q^2=0$), threshold
($q^2=9m^2$), and pseudothreshold ($q^2=m^2$), the values of the diagram can be
found in closed form.
At any other value of $q^2$, we have formal integral representations.
There is, however, another interesting point where one would like to have some
closed analytical expression, namely at $q^2=-m^2$.
Such kinematics occurs, for example, in the calculations of hard Wilson
coefficients in non-relativistic QED and QCD (see, e.g.,
Ref.~\cite{Czarnecki:2001gi}) at the next-to-leading order (NLO).
This can be applied, e.g., to evaluation of the parapositronium decay rate in
non-relativistic QED.
In non-relativistic QCD, applications are related to the analysis of
heavy-quarkonium decays and the near-threshold $t\bar{t}$ production at a
future $e^+e^-$ linear collider, which offers the opportunity to improve the
accuracy of the $t$-quark mass.


%
%
\begin {figure} [htbp]
\centerline{
\begin{picture}(150,100)(0,0)
\SetWidth{1.0}
\CArc(75,50)(30,0,180)
\Line(45,50)(25,50)
\Line(105,50)(125,50)
\SetWidth{3.0}
\CArc(75,50)(30,180,360)
\Line(45,50)(105,50)
\Text(75,30)[c]{$\scriptstyle c,\,M^2$}
\Text(75,60)[c]{$\scriptstyle b,\,M^2$}
\Text(75,87)[c]{$\scriptstyle a,\,m^2$}
\Text(115,60)[l]{$q^2=-m^2$}
\Text(75,5)[c]{(a)}
\end{picture}
~~~~~~~~~~~
\begin{picture}(150,100)(0,0)
\SetWidth{1.0}
\CArc(75,50)(30,0,180)
\Line(45,50)(25,50)
\Line(105,50)(125,50)
\SetWidth{3.0}
\CArc(75,50)(30,180,360)
\Text(75,30)[c]{$\scriptstyle b,\,M^2$}
\Text(75,87)[c]{$\scriptstyle a,\,m^2$}
\Text(115,60)[l]{$q^2=-m^2$}
\Text(75,5)[c]{(b)}
\end{picture}
}
\caption{(a) Two-loop sunset diagram $J_{abc}^{mMM}$ and (b) one-loop
self-energy diagram $I_{ab}^{mM}$ with different masses, $m$ and $M$.
In both cases, the external four-momentum $q$ satisfies the condition
$q^2=-m^2$.
The label attached to a given line indicates the mass square appearing in the
respective propagator and the power to which the latter is raised.}
\label{sunsetMMm}
\end{figure}
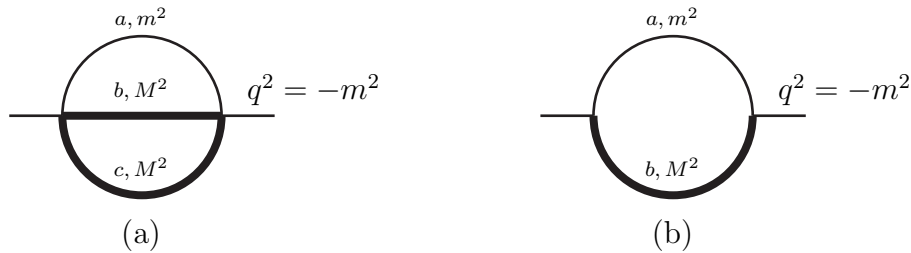

The goal of this paper is to fill this gap by finding a suitable
representation for the above-mentioned object.
Actually, we solve an even more general problem.
In fact, we evaluate the sunset diagram with two different masses, $m$ and
$M$, at $q^2=-m^2$, as shown in Fig.~\ref{sunsetMMm}(a).
The case of interest then simply emerges by setting $m=M$.

This paper is organized as follows.
In Section~\ref{sec:two}, we establish a relation between the two-loop sunset
diagram of Fig.~\ref{sunsetMMm}(a) and the one-loop one of
Fig.~\ref{sunsetMMm}(b).
In Section~\ref{sec:three}, we work out the one-loop case.
In Section~\ref{sec:four}, we consider the two-loop case for the general case
of $m\ne M$.
In Section~\ref{sec:five}, we derive a closed expression for the special case
of $m= M$.
Our conclusions are contained in Section~\ref{sec:six}.

\section{Relation between one- and two-loop sunset diagrams}
\label{sec:two}

In our calculation, we use a particular formula to represent a loop with two
massive propagators as an integral whose integrand contains a new propagator
with a mass that depends on the variable of integration.
Graphically, this formula has the following form:
\begin{eqnarray}
\mbox{{
\begin{picture}(60,30)(0,13)
\SetWidth{1.0}
\SetWidth{2.0}
\Curve{(5,15)(30,25)(55,15)}
\Curve{(5,15)(30,5)(55,15)}
\SetWidth{1.0}
\Line(5,15)(-5,15)
\Line(55,15)(65,15)
\Text(30,27)[b]{$\scriptstyle a,\,M^2$}
\Text(30,3)[t]{$\scriptstyle b,\,M^2$}
\Text(0,12)[t]{$q$}
\end{picture}
}}
\hspace{2mm}
&=&i^{1+d} \frac{\Gamma(a+b-d/2)}{\Gamma(a)\,\Gamma(b)}
   \int\limits_0^1 \frac{{\rm d}s}{(1-s)^{a+1-d/2}s^{b+1-d/2}}
\nonumber\\
&&{}\times\hspace{3mm}
\raisebox{1mm}{{
\begin{picture}(70,30)(0,4)
\SetWidth{2.0}
\Line(5,5)(65,5)
\Vertex(5,5){2}
\SetWidth{1.0}
\Vertex(65,5){2}
\Line(5,5)(-5,5)
\Line(65,5)(75,5)
\Text(33,7)[b]{$\scriptstyle a+b-d/2$}
\Text(33,7)[b]{}
\Text(33,1)[t]{$\scriptstyle M^2/[s(1-s)]$}
\Text(-3,-5)[b]{$q$}
\end{picture}
}}
\hspace{3mm},
\label{master}
\end{eqnarray}
where the loop with two propagators with mass square $M^2$ is replaced by one
propagator with mass square $M^2/[s(1-s)]$.
Here and in the following, we employ dimensional regularization in
$d=4-2\varepsilon$ space-time dimensions.

Equation~(\ref{master}) is easily derived from the Feynman parameter
representation and was introduced in Euclidean and Minkowski spaces in
Refs.~\cite{Fleischer:1997bw} and \cite{Fleischer:1998nb}, respectively.
Here, we work in Minkowski space and will, thus, follow
Ref.~\cite{Fleischer:1998nb}.

We shall adopt the following strategy: applying Eq.~(\ref{master}), we shall
represent the results for the sunset diagrams as integrals of one-loop
two-point diagrams involving propagators with masses that depend on the
variable of integration.
We shall first evaluate the one-loop integrals in Section~\ref{sec:three} and
then reconstruct the results for the two-loop sunset diagrams involving two
different masses with the help of Eq.~(\ref{master}) in Section~\ref{sec:four}.
A similar strategy was recently also adopted for the calculation of certain 
four-loop tadpole diagrams \cite{Kniehl:2005yc}.

\section{One-loop case}
\label{sec:three}

Let us first consider the one-loop case shown in Fig.~\ref{sunsetMMm}(b).
The self-energy integral in $d$ dimensions is given by
\begin{equation}
I_{ab}^{mM} = \int \frac{{\rm d}^dk}{\pi^{d/2}}\,
  \frac{1}{(k^2-m^2)^a[(k-q)^2-M^2]^b} \,,
\end{equation}
where $q^2=-m^2$ is implied.

First we try to establish the general structure of the result.
For this purpose it is convenient to consider the 
integral $I_{12}^{mM}$ rather than $I_{11}^{mM}$.
This integral is ultraviolet finite, so that we may put $d=4$.
(We will return to the case $d=4-2\varepsilon$ later on.)
Using the standard Feynman parameter technique, 
it is not difficult to obtain the representation
\begin{eqnarray}
\label{Iab}
I_{12}^{mM} = -i \int\limits_0^1
    \frac{{\rm d}t\,(1-t)}{(1-t)M^2 + t(2-t) m^2}\,,
\end{eqnarray}
which could be further evaluated in terms of logarithms.

Having in mind the algorithm proposed above, our next step will be to replace
the mass $M^2$ by $M^2/[s(1-s)]$ with $s$ integrated from 0 to 1.
The easiest way to perform such an integration is to expand Eq.~(\ref{Iab}) in
powers of $M^2$ or $1/M^2$.
Then, the integration over $s$ can be performed term by term leading to Euler
gamma functions.
We shall follow this logic and expand $I_{12}^{mM}$ in $1/M^2$.
The expansion in $M^2$, corresponding to another regime, is also possible,
but more involved and will not be considered here.

It turns out that Eq.~(\ref{Iab}) is not appropriate for obtaining the
coefficients of the expansion.
It is much more advantageous to start from a differential equation.
First, a differential equation for $I_{11}^{mM}$ is obtained through
integration by parts (for a review of this procedure, see
Refs.~\cite{Kotikov:1990kg,Kotikov:1991hm}).
Next, differentiating both sides of this equation w.r.t.\ $M^2$, one obtains
a differential equation for $I_{12}^{mM}$, exploiting that
$I_{12}^{mM}=({\rm d}/{\rm d}M^2)I_{11}^{mM}$.
This equation reads
\begin{eqnarray}
\left[
d-4 + \frac{4m^4}{M^4} - \left(1+\frac{4m^4}{M^4} \right)
 M^2 \frac{\rm d}{{\rm d}M^2}
\right] I_{12}^{mM} =  f_{12}\,,
\label{di4}
\end{eqnarray}
where $f_{12}$ is a non-uniform term which contains only vacuum tadpoles and
is given by
\begin{eqnarray}
f_{12}\left( \frac{m^2}{M^2} \right) =
    \frac{i}{m^2}\left[ - \frac{m^2}{M^2} - 2\frac{m^4}{M^4}
       \left(\ln\frac{m^2}{M^2} + 1\right) \right] + O(\varepsilon)\,.
\label{di5}
\end{eqnarray}
In order to solve the differential equation (\ref{di4}), we need one boundary
condition.
A suitable condition follows from the observation that $I_{12}^{mM}\to0$ as
$M^2\to \infty$.
Then, the solution of Eq.~(\ref{di4}) reads
\begin{eqnarray}
\frac{m^2}{i} I_{12}^{mM}(x) &=& \frac{1}{\sqrt{1+4x^2}}
    \int\limits^x_0
\frac{{\rm d}x'}{x'\sqrt{1+4x'^2}} \, \frac{m^2}{i}f_{12}(x') \nonumber \\
   &=& \frac{-1}{\sqrt{1+4x^2}} \,
   \int\limits^x_0  \frac{{\rm d}x'}{\sqrt{1+4x'^2}}
   ( 1 + 2x' + 2x'\ln x') \,,
\label{di6}
\end{eqnarray}
where $d=4$ and we have introduced the variable
\begin{equation}
\label{x}
  x = \frac{m^2}{M^2} \,.
\end{equation}
The first term in Eq.~(\ref{di6}) can be rewritten as
\begin{equation}
 \frac{-1}{\sqrt{1+4x^2}} \int\limits^x_0
  \frac{{\rm d}x'}{\sqrt{1+4x'^2}}  = \frac{-x}{\sqrt{1+4x^2}}
\,{}_2F_1\left(\frac{1}{2},\frac{1}{2};\frac{3}{2};-4x^2\right) \, ,
\label{di7}
\end{equation}
where ${}_2F_1$ is the hypergeometric function \cite{Abramowitz}.
Using the property of the hypergeometric function \cite{Abramowitz}
\begin{equation}
{}_2F_1\left(a,b;c;x\right) = (1-x)^{c-a-b}
\,{}_2F_1\left(c-a,c-b;c;x\right)\,,
\label{di8}
\end{equation}
Eq.~(\ref{di7}) can be transformed into
\begin{equation}
  - x \,{}_2F_1\left(1,1;\frac{3}{2};-4x^2\right) =
   \frac{1}{x} \sum_{n=1}^{\infty}
      (-16x^2)^n \frac{1}{\left( {2n\atop n}\right)}\,
          \frac{1}{8n} \,.
\end{equation}
The sum of the last two terms on the r.h.s.\ of Eq.~(\ref{di6}) yields
\begin{eqnarray}
\label{sum}
-\frac{2}{\sqrt{1+4x^2}}\int\limits^x_0
  \frac{{\rm d}x'\, x'}{\sqrt{1+4x'^2}}(\ln x' +1)
&=& \frac{1}{2} \left( \frac{1}{\sqrt{1+4x^2}} - 1 \right)
     \ln x
\\
&&{}   - \frac{1}{4\sqrt{1+4x^2}}
\sum_{n=1}^{\infty} \frac{\Gamma(n+1/2)}{n!\Gamma(1/2)}\, \frac{(-4x^2)^n}{n}
\,.
\nonumber
\end{eqnarray}
The term in the second line of Eq.~(\ref{sum}) can be represented as a limit,
\begin{equation}
\lim_{\delta\to0}\frac{1}{\delta}\left[
{}_2F_1\left(\frac{1}{2},\delta;1+\delta;-4x^2\right) -1 \right] \,.
\nonumber
\end{equation}
Now, using property (\ref{di8}), we can remove the square root
$\sqrt{1+4x^2}$.
Finally, after some transformations, we obtain the following series
representation for the one-loop diagram $I_{12}^{mM}$ with $d=4$:
\begin{eqnarray}
\frac{m^2}{i} I_{12}^{mM}(x) &=&
\sum_{n=1}^{\infty}
      (-x^2)^n \left( {2n\atop n}\right)
     \left( \frac{1}{2}\ln x
      - \frac{1}{2}S_1(n-1)
      + \frac{1}{2}S_1(2n-1)
      - \frac{1}{4n} \right)  \nonumber\\
    && +
 \frac{1}{x} \sum_{n=1}^{\infty}
      (-16x^2)^n \frac{1}{\left( {2n\atop n}\right)}\,
          \frac{1}{8n} \,,
\label{I12series}
\end{eqnarray}
where $S_1(n)=\sum_{j=1}^n 1/j$ is the harmonic sum.
This is just the form we wanted to have.
The appearance of the central binomial coefficient here is not surprising.
It corresponds to the branch point at $x=4$.
Similar one- and two-loop diagrams were considered in
Ref.~\cite{Jegerlehner:2002em}, and binomial sums were studied in
Ref.~\cite{Kalmykov:2000qe}.

Now, with the knowledge of the generic $n$-th coefficient of the expansion in
$x$, it is not difficult to find subsequent terms in the $\varepsilon$
expansion.
To do this, one has to increase the ``weights'' of the harmonic sums arising
from $1/n$, $S_1$ to $1/n^2$, $S_1^2$, and $S_2$ and so on.
At $O(\varepsilon)$, we thus obtain
\begin{eqnarray}
 \frac{m^2}{i} I_{12}^{mM,O(\varepsilon)}(x) &=&
     \varepsilon \sum_{n=1}^{\infty}
      (-x^2)^n \left( {2n\atop n}\right)
     \left[ \frac{1}{4}\ln^2 x
          + \left( \frac{1}{2}S_1(2n-1) + \frac{1}{4n} \right) \ln x \right.
 \nonumber\\
   &&{} - \frac{1}{8} S_2(n-1)
        - \frac{1}{4} S_2(2n-1)
        - \frac{1}{4} S_1^2(n-1)  \nonumber\\
     &&{}+ \left.
         \frac{1}{4} S_1^2(2n-1)
        - \frac{1}{2} \frac{S_1(n-1)}{n}
        + \frac{1}{4} \frac{S_1(2n-1)}{n}
        - \frac{3}{8n^2} \right] \nonumber\\
    &&{} +
  \varepsilon\frac{1}{x} \sum_{n=1}^{\infty}
      (-16x^2)^n \frac{1}{\left( {2n\atop n}\right)}
       \left( \frac{1}{8n} \ln x
         + \frac{S_1(2n-1)}{8n} \right) \,.
\end{eqnarray}



\section{Two-loop sunset diagram}
\label{sec:four}

Now, we are ready to evaluate the two-loop sunset diagram $J_{abc}^{mMM}$
shown in Fig.~\ref{sunsetMMm}(a).
Its loop-integral representation reads
\begin{equation}
J_{abc}^{mMM} = \int \frac{{\rm d}^dk\,{\rm d}^dl}{\pi^d}
  \frac{1}{(k^2-m^2)^a(l^2-M^2)^b[(l-k-q)^2-M^2]^c} \,,
\end{equation}
where $q^2=-m^2$ is implied.

Exploiting again Eq.~(\ref{master}), we replace $M^2$ in the one-loop
subdiagram by $M^2/[s(1-s)]$ and integrate over $s$.
It is known, e.g.\ from Ref.~\cite{Tarasov:1997kx}, that for such kinematics
there exist three master-integrals.
As master integrals we choose $J_{111}^{mMM}$, $J_{112}^{mMM}$, and
$J_{122}^{mMM}$.

After some calculation, we finally obtain
\begin{eqnarray}
M^2 J_{122}^{mMM}&=&
    \frac{1}{x} \sum_{n=1}^{\infty} (-x^2)^n
    \frac{\left(2n \atop n\right)}{\left(4n \atop 2n\right)} \left(
       - \frac{1}{2n} \ln x
       + \frac{S_1}{2n}
       - \frac{3}{2} \frac{\overline{S_1}}{n}
       + \frac{\overline{\overline{S_1}}}{n}
       + \frac{1}{4n^2}
       \right)   \nonumber\\
&&{}   + \frac{1}{x^2} \sum_{n=1}^{\infty}
     \frac{(-16x^2)^n}{\left(2n \atop n\right)\left(4n \atop 2n\right)}
      \left( - \frac{1}{2n-1} + \frac{1}{2n} - \frac{1}{4n^2}  \right) \,,
\label{J122}\\
%
%
(M^2)^{2\varepsilon} J_{112}^{mMM} &=&
     - \frac{1}{2\varepsilon^2} - \frac{1}{2\varepsilon}
    - \frac{1}{\varepsilon} \ln x  - \ln^2 x - \ln x
   - \frac{1}{2}\zeta(2) - \frac{1}{2}
               \nonumber\\
&&{}   + \frac{1}{x} \sum_{n=1}^{\infty} (-x^2)^n
    \frac{\left(2n \atop n\right)}{\left(4n \atop 2n\right)}\,
        \frac{4n-1}{n(2n-1)^2}   \left(
         \frac12 \ln x - \frac{1}{2} S_1
           + \frac{3}{2} \overline{S_1} - \overline{\overline{S_1}}
                  \right. \nonumber\\
&&{}- \left.
           \frac{1}{4n} - \frac{1}{2n-1} + \frac{1}{4n-1}
       \right)
+ \sum_{n=1}^{\infty}
     \frac{(-16x^2)^n}{\left(2n \atop n\right)\left(4n \atop 2n\right)}\,
      \frac{-1}{4n^2(2n+1)}  \,,
\label{J112}\\
%
%
(M^2)^{-1+2\varepsilon} J_{111}^{mMM} &=&
     - \frac{1}{\varepsilon^2}  - \frac{3}{\varepsilon}
   - \frac{2}{\varepsilon} \ln x - 2 \ln^2 x - 6 \ln x
      - \zeta(2) - 7
         \nonumber\\
&&{} + x \left( - \frac{1}{2\varepsilon^2} - \frac{7}{4\varepsilon}
        + \frac{1}{2} \ln^2 x - \frac{5}{2} \ln x
        - \frac{1}{2} \zeta(2) + \frac{5}{8}   \right)
             \nonumber\\
&&{} + x \sum_{n=1}^{\infty} (-x^2)^n
    \frac{\left(2n \atop n\right)}{\left(4n \atop 2n\right)}\,
        \frac{1}{n(n+1)(4n+1)}   \left(
            \frac{1}{2} \ln x
           - \frac{1}{2} S_1 + \frac{3}{2} \overline{S_1}
              - \overline{\overline{S_1}}  \right. \nonumber\\
&&{}-   \left. \frac{1}{4n} - \frac{1}{4(n+1)} - \frac{1}{4n+1}
       \right)   \nonumber\\
&&{} + \sum_{n=1}^{\infty}
     \frac{(-16x^2)^n}{\left(2n \atop n\right)\left(4n \atop 2n\right)}\,
      \frac{1}{2n^2(2n+1)(2n-1)}  \,,
\label{J111}
\end{eqnarray}
where $\zeta$ denotes Riemann's zeta function and, for brevity, we omitted the
arguments of the harmonic sum $S_1(n)$ and introduced the short-hand notations
\begin{eqnarray}
S_1 = S_1(n-1)\,, \qquad \overline{S_1} = S_1(2n-1)\,,
   \qquad \overline{\overline{S_1}} = S_1(4n-1) \,.
\end{eqnarray}

For the practical applications mentioned in the Introduction,
we also need the $O(\varepsilon)$ terms of the sunset diagrams.
We do not present them here for a general value of $x$, since the
corresponding expressions are cumbersome.
However, we shall do this in the next sections for the equal-mass case, which
is of most interest.


\section{Equal-mass case}
\label{sec:five}

We now turn to the case when $m=M$.
Note that, with equal masses, there are only two master integrals,
$J_{111}$ and $J_{112}$.
After substituting $x=1$ in Eqs.~(\ref{J122})--(\ref{J111}), one obtains
series which cannot be summed explicitly to become some known constants.
However, they can be slightly simplified.
In fact, using the PSLQ algorithm \cite{PSLQ}, which is able to reconstruct
the rational-number coefficients multiplying a given set of irrational numbers
from a high-precision numerical result, we find that there are some relations
among sums at $x=1$ and the irrationals $\ln2$ and $\zeta(2)$.
Specifically, we can establish the following.
Among the sums occurring in Eqs.~(\ref{J122})--(\ref{J111}),
\begin{eqnarray}
\lefteqn{\sum_{n=1}^{\infty} (-1)^n
    \frac{\left(2n \atop n\right)}{\left(4n \atop 2n\right)}
  \left\{  \frac{1}{n},\,\frac{1}{n^2},\,
          \frac{1}{n+1},\,\frac{1}{(n+1)^2},\,
          \frac{1}{2n-1},\,\frac{1}{(2n-1)^2},\,
          \frac{1}{4n-1},\right.}\nonumber\\
 &&{} \left.
          \frac{1}{4n+1},\,\frac{1}{(4n+1)^2},\,
          \frac{\phi}{n},\,\frac{\phi}{n+1},\,
          \frac{\phi}{2n-1},\,\frac{\phi}{(2n-1)^2},\,
          \frac{\phi}{4n+1}
       \right\} \,, \nonumber
\end{eqnarray}
where $\phi=S_1-3\overline{S_1}+2\overline{\overline{S_1}}$,
are not independent.
To this list, we add also the corresponding sums with the prefactors $1$ or
$\phi/n^2$.
All of them can be expressed in terms of five sums and the constants $\ln2$
and $\zeta(2)$.
We are able to evaluate two of these sums in terms of known functions.
Let us introduce the notation
\begin{equation}
  p = \sqrt{5}, \qquad \varphi = \arctan\sqrt{p} \,.
\end{equation}
Then, the new constants read
\begin{eqnarray}
  f_1 &=& \frac{1}{\sqrt{p}}
   F\left( 2 \arctan\sqrt{p},\, \frac{1+p}{2p}\right)
= \frac{1}{\sqrt{p}}
   F\left( 2 \varphi ,\, \frac{1}{2\sin^2\varphi}\right)
   = 1.8829167613\dots \,,\\
  e_1 &=& \frac{1}{\sqrt{p}}
   E\left( 2 \arctan\sqrt{p},\, \frac{1+p}{2p}\right)
= \frac{1}{\sqrt{p}}
   E\left( 2 \varphi ,\, \frac{1}{2\sin^2\varphi}\right)
   = 0.9671227369\dots  \,,
\end{eqnarray}
where $F$ and $E$ are the elliptic functions of the first and second
kind, respectively.
For the reader's convenience, we recall here their definitions:
\begin{equation}
F(\phi,k)=\int_0^\phi{\rm d}\theta\, (1-k\sin^2\theta)^{-1/2}\,,
\qquad
E(\phi,k)=\int_0^\phi{\rm d}\theta\, (1-k\sin^2\theta)^{1/2}\,.
\end{equation}
With these constants, we evaluate the following two sums:
\begin{eqnarray}
  \sum_{n=1}^{\infty} (-1)^n
    \frac{\left(2n \atop n\right)}{\left(4n \atop 2n\right)} 
   &=&  -\frac{1}{2} + \frac{1}{10}p
     + \frac{1}{5}p e_1 - \frac{1}{10}p f_1 \,,
\label{a0}\\
  \sum_{n=1}^{\infty} \frac{(-1)^n}{n}\,
    \frac{\left(2n \atop n\right)}{\left(4n \atop 2n\right)}
      &=& - \frac{4}{3}\ln2 +\frac{1}{3}f_1 \,.
\label{a1}
\end{eqnarray}
It seems that other possible irrationalities will include some integrals of
elliptic functions.
This requires further investigations.

Among the five sums of the other kind appearing in
Eqs.~(\ref{J122})--(\ref{J111}),
\begin{equation}
  \sum_{n=1}^{\infty}
  \frac{(-16)^n}{\left(2n \atop n\right)\left(4n \atop 2n\right)}
  \left\{ 1,\, \frac{1}{n},\,\frac{1}{n^2},\,
          \frac{1}{(2n-1)},\,\frac{1}{(2n+1)} \right\}\,,
\end{equation}
only two are independent.
Here is an example of a relation between these sums:
\begin{equation}
  \sum_{n=1}^{\infty}
  \frac{(-16)^n}{\left(2n \atop n\right)\left(4n \atop 2n\right)}
    \left( 20 - \frac{16}{n} + \frac{3}{n^2} \right) = -4\,.
\end{equation}
It is interesting to observe that sums of different ``weights,'' namely 1,
$1/n$, and $1/n^2$, appear here.
In earlier studies, the independence of sums with different ``weights'' was
encountered and frequently used (see, e.g.,
Refs.~\cite{Fleischer:1997bw,Fleischer:1998nb,Fleischer:1999hp}).

A similar analysis can also be performed at $O(\varepsilon)$.
Finally, setting $m=M=1$, we obtain for the master integrals through
$O(\varepsilon)$
\begin{eqnarray}
J_{111}^{mmm} &=& -\frac{3}{2\varepsilon^2} - \frac{19}{4\varepsilon}
   + \Sigma_{111} + \varepsilon \Sigma_{111}^{O(\varepsilon)}
   + O(\varepsilon^2) \,,\\
J_{112}^{mmm} &=& -\frac{1}{2\varepsilon^2} - \frac{1}{2\varepsilon}
   + \Sigma_{112} + \varepsilon \Sigma_{112}^{O(\varepsilon)} 
   + O(\varepsilon^2) \,,
\end{eqnarray}
where
\begin{eqnarray}
  \Sigma_{111} & = &     
 - \frac{215}{24}  + \frac{9}{4}\zeta(2) \nonumber\\
  &&{} + \sum_{n=1}^{\infty} (-1)^n
    \frac{\left(2n \atop n\right)}{\left(4n \atop 2n\right)}
    \left( - \frac{5}{2}\phi + \frac{15\phi}{4n} + \frac{15}{8n^2}
       \right) 
     + \sum_{n=1}^{\infty}
     \frac{(-16)^n}{\left(2n \atop n\right)\left(4n \atop 2n\right)}
     \left( \frac{25}{3} - \frac{25}{6n} \right)  
      \nonumber\\
     & = & -9.03056576107922587907436223954936770213033473413\dots \,, \\
  \Sigma_{112} & = &  
      \frac{29}{18} + \frac{3}{4}\zeta(2) \nonumber\\
 &&{} + \sum_{n=1}^{\infty} (-1)^n
    \frac{\left(2n \atop n\right)}{\left(4n \atop 2n\right)}
    \left( - \frac{5}{2}\phi + \frac{5\phi}{4n} + \frac{5}{8n^2}
       \right)  
     + \sum_{n=1}^{\infty}
     \frac{(-16)^n}{\left(2n \atop n\right)\left(4n \atop 2n\right)}
     \left( \frac{50}{9} - \frac{35}{18n} \right)
  \nonumber\\
   & = & 0.0113804720812563731826135489564394881100859890024139\dots \,,
\nonumber\\
\Sigma_{111}^{O(\varepsilon)} &=&
  \zeta(3) + \frac{49}{8}\zeta(2) - \frac{15}{2}\zeta(2)\ln2 
     - \frac{1607}{144} \nonumber\\
  &&{} + \sum_{n=1}^\infty (-1)^n 
    \frac{\left(2n\atop n\right)}{\left(4n\atop 2n\right)} 
   \left(
      -\frac{65}{4} \phi 
     - \frac{5}{8} \sigma
     - \frac{5}{4} \phi\rho
     + \frac{77}{8}\, \frac{\phi}{n}
     + \frac{15}{16}\, \frac{\sigma}{n}
     + \frac{15}{8}\, \frac{\phi\rho}{n}
     + \frac{77}{16n^2} 
     + \frac{15}{4}\, \frac{\kappa}{n^2}
\right.\nonumber\\
&&{}+\left. \frac{45}{16n^3} 
       \right)
   + \sum_{n=1}^\infty 
     \frac{(-16)^n}{\left(2n\atop n\right)\left(4n\atop 2n\right)} 
   \left(
     \frac{655}{18} 
     + \frac{50}{3} \beta
     - \frac{493}{36n} 
     - \frac{25}{3} \frac{\beta}{n}
      \right) \nonumber\\
  &=& -25.4473853869869254012665862897193562434443269569232788372\dots
      \,, \\
\Sigma_{112}^{O(\varepsilon)} &=&
  \frac{1}{3}\zeta(3) + \frac{7}{12}\zeta(2) - \frac{5}{2}\zeta_2\ln2 
     + \frac{485}{54} \nonumber\\
  &&{} + \sum_{n=1}^\infty (-1)^n 
    \frac{\left(2n\atop n\right)}{\left(4n\atop 2n\right)} 
   \left(
      -\frac{25}{2} \phi 
     - \frac{5}{8} \sigma
     - \frac{5}{4} \phi\rho
     - \frac{\phi}{n}
     + \frac{5}{16}\, \frac{\sigma}{n}
     + \frac{5}{8}\, \frac{\phi\rho}{n}
     + \frac{5}{4}\, \frac{\kappa^2}{n}
     - \frac{1}{2n^2}
\right.\nonumber\\ 
 &&{}+\left. \frac{15}{16n^3} 
       \right)
  + \sum_{n=1}^\infty 
     \frac{(-16)^n}{\left(2n\atop n\right)\left(4n\atop 2n\right)} 
   \left(
     \frac{470}{27} 
     + \frac{100}{9} \beta
     - \frac{107}{54n} 
     - \frac{35}{9}\, \frac{\beta}{n}
      \right)  \nonumber\\
  &=&  -2.04235726589417810442922843329658331200829586895\dots  \,.
\end{eqnarray}
Here, we have introduced the following short-hand notations
\begin{eqnarray}
\phi &=& S_1(n-1) - 3S_1(2n-1) + 2S_1(4n-1) \,, \nonumber\\
\rho &=& S_1(n-1) + 5S_1(2n-1) - 2S_1(4n-1) \,, \nonumber\\
\sigma &=& S_2(n-1) + 10S_2(2n-1) - 8S_2(4n-1) \,, \nonumber\\
\kappa &=& S_1(n-1) - S_1(2n-1) + S_1(4n-1) \,, \nonumber\\
\beta &=&  2S_1(2n-1) - S_1(4n-1) \,.
\end{eqnarray}


\section{Conclusions}
\label{sec:six}

In this paper, we considered the two-loop sunset diagram with two different
masses, $m$ and $M$, in the special kinematical regime where $q^2 = -m^2$.
The coefficients of the expansion in the variable $x=m^2/M^2$ were found
explicitly.
This allows for the numerical restoration of the result, for example via
Pad\'e approximants.
In the special case of $m=M$, the result was expressed in terms of
very fast-converging alternating series.
Some analytical relations between these series were found using the PSLQ
algorithm \cite{PSLQ}.
Their exact analytical structure is related to elliptic integrals.
The two lowest elements of this structure were expressed in terms of the
elliptic functions $E$ and $F$.
Integral representations for the sunset master integrals $J_{111}^{mMM}$, 
$J_{112}^{mMM}$, and $J_{122}^{mMM}$ will be presented in a future
publication.
\bigskip

\noindent
{\bf Acknowledgments}

\smallskip

The authors are grateful to M.Yu.\ Kalmykov for fruitful discussions and
valuable comments.
The work of A.V.K. and O.V. was supported in part by RFBR Grant No.\
N~05-02-17645.
The work of A.V.K. was also supported in part by the
Heisenberg-Landau-Programm.
This work was supported in part by BMBF Grant No.\ 05~HT4GUA/4 and HGF Grant
No.\ NG-VH-008.


\begin{thebibliography}{99}

\bibitem{sunsetpub}
  P.~Post, J.B.~Tausk,
  Mod.\ Phys.\ Lett.\ A 11 (1996) 2115;\\
  F.A.~Berends, A.I.~Davydychev, N.I.~Ussyukina,
  Phys.\ Lett.\ B 426 (1998) 95;\\
 S.~Groote, J.G.~Korner, A.A.~Pivovarov,
  Phys.\ Lett.\ B 443 (1998) 269;\\
 S.~Groote, J.G.~Korner, A.A.~Pivovarov,
  Nucl.\ Phys.\ B 542 (1999) 515;\\
  S.~Groote, J.G.~Korner, A.A.~Pivovarov,
  Eur.\ Phys.\ J.\ C 11 (1999) 279;\\
  A.I.~Davydychev, V.A.~Smirnov,
  Nucl.\ Phys.\ B 554 (1999) 391;\\
  A.~Bashir, R.~Delbourgo, M.L.~Roberts,
  J.\ Math.\ Phys.\ 42 (2001) 5553;\\
  P.~Mastrolia, E.~Remiddi,
  Nucl.\ Phys.\ B 657 (2003) 397;\\
  A.~Onishchenko, O.~Veretin,
Yad.\ Fiz.\  68 (2005) 1461
[Phys.\ Atom.\ Nucl.\ 68 (2005) 1405].

\bibitem{Kotikov:1990kg}
  A.V.~Kotikov,
  Phys.\ Lett.\ B 254 (1991) 158;\\
  A.V.~Kotikov,
  Mod.\ Phys.\ Lett.\ A 6 (1991) 677.

\bibitem{Fleischer:1998dw}
  J.~Fleischer, F.~Jegerlehner, O.V.~Tarasov, O.L.~Veretin,
  Nucl.\ Phys.\ B 539 (1999) 671;\\
  J.~Fleischer, F.~Jegerlehner, O.V.~Tarasov, O.L.~Veretin,
  Nucl.\ Phys.\ B 571 (2000) 511, Erratum;\\
  A.I.~Davydychev, M.Yu.~Kalmykov,
  Nucl.\ Phys.\ B 605 (2001) 266;\\
  F.~Jegerlehner, M.Yu.~Kalmykov, O.~Veretin,
  Nucl.\ Phys.\ B 658 (2003) 49.

\bibitem{Berends:1993ee}
  F.A.~Berends, M.~Buza, M.~B\"ohm, R.~Scharf,
  Z.\ Phys.\ C 63 (1994) 227;\\
  S.~Bauberger, F.A.~Berends, M.~B\"ohm, M.~Buza,
  Nucl.\ Phys.\ B 434 (1995) 383.

\bibitem{Davydychev:2003cw}
  D.J.~Broadhurst,
  Z.\ Phys.\ C 47 (1990) 115;\\
  A.I.~Davydychev, R.~Delbourgo,
  J.\ Phys.\ A 37 (2004) 4871.

\bibitem{Broadhurst:1993mw}
  D.J.~Broadhurst, J.~Fleischer, O.V.~Tarasov,
  Z. Phys.\ C 60 (1993) 287.

\bibitem{Laporta:2004rb}
  S.~Laporta, E.~Remiddi,
  Nucl.\ Phys.\ B 704 (2005) 349.

\bibitem{Czarnecki:2001gi}
  A.~Czarnecki, K.~Melnikov, A.~Yelkhovsky,
  Phys.\ Rev.\ A 61 (2000) 052502;\\
  A.~Czarnecki, K.~Melnikov,
  Phys.\ Rev.\ D 65 (2002) 051501.

\bibitem{Fleischer:1997bw}
  J.~Fleischer, A.V.~Kotikov, O.L.~Veretin,
  Phys.\ Lett.\ B 417 (1998) 163.

\bibitem{Fleischer:1998nb}
  J.~Fleischer, A.V.~Kotikov, O.L.~Veretin,
  Nucl.\ Phys.\ B 547 (1999) 343.

\bibitem{Kniehl:2005yc}
  B.A.~Kniehl, A.V.~Kotikov,
 Report No.\ DESY~05-153, hep-ph/0508238.

\bibitem{Kotikov:1991hm}
  A.V.~Kotikov,
  Phys.\ Lett.\ B 259 (1991) 314;\\
  A.V.~Kotikov,
  Phys.\ Lett.\ B 267 (1991) 123;\\
  E. Remiddi,
  Nuovo Cim.\ A 110 (1997) 1435.


\bibitem{Abramowitz}
M. Abramowitz, I.A. Stegun (Eds.), Handbook of Mathematical Functions with
Formulas, Graphs, and Mathematical Tables, National Bureau of Standards
Applied Mathematics Series, vol.\ 55, US Government Printing Office,
Washington, DC, 1964.

\bibitem{Jegerlehner:2002em}
  F.~Jegerlehner, M.Yu.~Kalmykov, O.~Veretin,
  Nucl.\ Phys.\ B 641 (2002) 285;\\
  F.~Jegerlehner, M.Yu.~Kalmykov, O.~Veretin,
  Nucl.\ Phys.\ B 658 (2003) 49.

\bibitem{Kalmykov:2000qe}
  M.Yu.~Kalmykov, O.~Veretin,
  Phys.\ Lett.\ B 483 (2000) 315;\\
  A.I.~Davydychev, M.Yu.~Kalmykov,
  Nucl.\ Phys.\ B 699 (2004) 3.

\bibitem{Tarasov:1997kx}
  O.V.~Tarasov,
  Nucl.\ Phys.\ B 502 (1997) 455.

\bibitem{PSLQ}
H.R.P.~Ferguson, D.H.~Bailey, RNR Technical Report No.\ RNR-91-032;\\
H.R.P.~Ferguson, D.H.~Bailey, S.~Arno, NASA Technical Report No.\ NAS-96-005.

\bibitem{Fleischer:1999hp}
  J.~Fleischer, M.Yu.~Kalmykov, A.V.~Kotikov,
  Phys.\ Lett.\ B 462 (1999) 169;\\
  A.V.~Kotikov, L.N. Lipatov,
  Nucl.\ Phys.\ B 582 (2000) 19;\\
 A.V.~Kotikov, L.N. Lipatov,
  Nucl.\ Phys.\ B 661 (2003) 19;\\
 A.V.~Kotikov, L.N. Lipatov, A.I. Onishchenko, V.N. Velizhanin,
  Phys.\ Lett.\ B 595 (2004) 521;\\
  Z.~Bern, L.J.~Dixon, D.A.~Kosower,
  Comptes Rendus Physique 5 (2004) 955;\\
  Z.~Bern, L.J.~Dixon, V.A.~Smirnov,
  Phys.\ Rev.\ D 72 (2005) 085001;\\
  S.~Moch, J.A.M.~Vermaseren, A.~Vogt,
JHEP 0508 (2005) 049;\\
  T. Gehrmann, T. Huber, D. Ma\^\i tre,
Phys.\ Lett.\ B 622 (2005) 295.

\end{thebibliography}
\end{document}